\documentclass[12pt,reqno]{amsart}  
\usepackage[left=2cm, right=2cm, top=2cm, bottom=2cm]{geometry}
\usepackage{amsmath,amsthm,amssymb,natbib,url,enumerate,tikz}
\usepackage[colorlinks,linkcolor=blue,citecolor=blue!50!black,urlcolor=blue]{hyperref}

\theoremstyle{definition}

\theoremstyle{remark}

\newcommand*{\set}[1]{\left\{#1\right\}}

\title{Corrigendum to ``Managerial Incentive Problems: A Dynamic Perspective''} 

\author{Sander Heinsalu
}
\thanks{Research School of Economics, Australian National University. HW Arndt Building, 25a Kingsley St, Acton ACT 2601, Australia. Email: sander.heinsalu@anu.edu.au, website: \url{https://sanderheinsalu.com/}
}
\date{\today}
\begin{document}
\maketitle
\begin{abstract}
This paper corrects some mathematical errors in \cite{holmstrom1999} and clarifies the assumptions that are sufficient for the results of \cite{holmstrom1999}. The results remain qualitatively the same. 

Keywords: Career concerns, symmetric incomplete information, dynamic games. 
	
JEL classification: C72, C73, D83. 
\end{abstract}

\section{Introduction}

This note corrects some mathematical errors in the career concerns model with normally distributed type and signal in \cite{holmstrom1999} Sections~2.1--2.3. 
The corrigendum is organized as follows. Section~\ref{sec:model} introduces the model and the additional assumptions on the cost function that are sufficient for some of the results in \cite{holmstrom1999}. 
Section~\ref{sec:basic} derives the equilibrium strategy in detail, correcting some errors in the derivation of \cite{holmstrom1999}. 
Section~\ref{sec:stationary} first corrects an inconsistency between \cite{holmstrom1999} Sections~2.1 and~2.2. Second, it introduces an assumption necessary and sufficient for the equilibrium labor supply to converge to zero as the type becomes persistent or the random component of the output large. Third, it discusses ways to resolve the indeterminacy of equilibrium in the undiscounted case. 
Section~\ref{sec:transient} presents two ways to modify the definitions in \cite{holmstrom1999} Section~2.3, both of which correct an error in the proof of Proposition~2.

\section{The model} 
\label{sec:model}

The notation and the model follow \cite{holmstrom1999}. 
The players are a manager and a competitive market. Time is discrete, indexed by $t\in\mathbb{N}$, and the horizon is infinite. 
The manager initially has ability $\eta_{1}$, which is symmetric incomplete information, and commonly believed to be drawn from the prior probability distribution $N(m_{1},1/h_{1})$, where $h_{1}\in(0,\infty)$ is the precision (the inverse of the variance). 
Ability evolves according to $\eta_{t+1}=\eta_{t}+\delta_{t}$, where $\delta_{t}\sim N(0,1/h_{\delta})$ i.i.d.\ over time. The basic model of \cite{holmstrom1999} assumes $h_{\delta}=\infty$, which is later relaxed. This corrigendum treats the cases $h_{\delta}=\infty$ and $h_{\delta}\in(0,\infty)$ together.

Each period $t$, the manager chooses labor $\hat{a}_{t}\in[0,\infty]$, which generates public output $y_{t}=\eta_{t}+\hat{a}_{t}+\epsilon_{t}$, where $\epsilon_{t}\sim N(0,1/h_{\epsilon})$ i.i.d.\ over time, $h_{\epsilon}\in(0,\infty)$, and $\epsilon_t$ is independent of $\eta_{1},\delta_s$ for any $t,s$.\footnote{
The independence of $\epsilon_t,\eta_{1},\delta_s$ for all $t,s$ is implicit in \cite{holmstrom1999}. 
} 
The competitive and risk-neutral market observes $y^{t-1}=(y_{1},\ldots,y_{t-1})\in\mathbb{R}^{t-1}$ at the start of period $t$ and pays the manager a wage $w_{t}(y^{t-1})\in[-\infty,\infty]$ equal to the manager's expected output. 
The market's wage rule $w=(w_{t})_{t=1}^{\infty}$ is a sequence of functions $w_{t}:\mathbb{R}^{t-1}\rightarrow[-\infty,\infty]$. 

The manager's cost of choosing labor $\hat{a}$ is $g(\hat{a})$, with $g$ continuously differentiable on $\mathbb{R}_{+}$, increasing, convex, $g'(0)=0$ and $\lim_{\hat{a}\rightarrow\infty}g'(\hat{a})=\infty$. 
\cite{holmstrom1999} only assumes that $g$ is increasing and convex. Sufficient for the maximizer of the manager's utility function~(\ref{utility}) to exist is that $g$ is continuous, because the action set $[0,\infty]$ is compact. 
For the first order approach used in \cite{holmstrom1999} to be valid, a convex and continuously differentiable $g$ is sufficient. 

The manager's public strategy $a=(a_{t})_{t=1}^{\infty}$ is a sequence of functions $a_t:\mathbb{R}^{t-1}\rightarrow[0,\infty]$, where $a_{t}$ maps the output history $y^{t-1}$ to the action $\hat{a}_{t}$. 
Public strategies (even mixed) are w.l.o.g.\ when the game has a product structure\footnote{A product structure means that the informativeness of the public signal about the actions of the other strategic players is independent of a player's own action.}, as shown in \cite{mailath+samuelson2006} p.~330. The game here has only one strategic player, so trivially a product structure. 
Hereafter, a public strategy is simply called strategy. 

The manager's discount factor is $\beta\in[0,1)$. The undiscounted model with $\beta=1$ is discussed at the end of Section~\ref{sec:stationary}. 
The manager's \emph{ex post} utility is 
\begin{align}
\label{utility}
U(w,a)=\sum_{t=1}^{\infty}\beta^{t-1}[w_{t}-g(a_{t})]. 
\end{align}

A \emph{perfect public equilibrium} consists of the manager's strategy $a^*$ and the market's wage rule $w$ such that for all $t$ and $y^{t-1}$, 
\begin{align}
\label{equilibrium}
&a_{t}^*(y^{t-1})\in\arg\max_{\hat{a}_{t}}\set{w_{t}(y^{t-1})-g(\hat{a}_{t}) + \sum_{\tau=t+1}^{\infty}\beta^{\tau-t}\mathbb{E}\left[w_{\tau}(y^{\tau-1})-g(a_{\tau}^{*}(y^{\tau-1}))\middle|y^{t-1}\right]},
\\&\notag 
w_{t}(y^{t-1})=\mathbb{E}\left[y_{t}|y^{t-1}\right] =\mathbb{E}\left[\eta_{t}|y^{t-1}\right]+a_{t}^*(y^{t-1}). 
\end{align}
Hereafter, a pure perfect public equilibrium is simply called equilibrium. 
Restricting attention to pure strategies is w.l.o.g., because $g$ is convex and any $\mathbb{E}[y_{t+k}|y^{t-1}]$ that the manager can generate with mixed actions can be generated by pure actions.

\subsection{Results for the basic model}
\label{sec:basic}

The market is Bayesian and conjectures the manager's strategy $a_{t}^*(y^{t-1})$, so de-biases the output $y_t$ to the signal $z_t :=y_t-a_t^*(y^{t-1})$, which in equilibrium equals $\eta_{t}+\epsilon_t $. From the manager's perspective, $z_{t} =\eta_{t}+\epsilon_{t}+\hat{a}_{t}-a_{t}^*(y^{t-1})$. 
The mean of the market's belief at the start of period $t$ (after $t-1$ signals) is denoted $m_t$ and the precision $h_{t}$. The precision of the market's belief after seeing $y_{t}$ but before taking into account the shock $\delta_{t}$ added to $\eta_{t}$ is $h_{t}+h_{\epsilon}$.\footnote{
To derive $h_{t+1}$, the market first updates $\eta_{t}$ (which has precision $h_{t}$) in response to the signal $z_{t}$ (precision $h_{\epsilon}$), obtaining precision $h_{t}+h_{\epsilon}$. Then the market adds the variance $\frac{1}{h_{\delta}}$ of the normally distributed shock $\delta_{t-1}$ to the variance $\frac{1}{h_{t-1}+h_{\epsilon}}$ of the normally distributed updated $\eta_{t-1}$. 
If $h_{\delta}=\infty$, then $h_{t}+h_{\epsilon}=h_{t+1}$ for all $t$. 

\cite{holmstrom1999} defines $\hat{h}_{t}:=h_{t}+h_{\epsilon}$, but interprets it on p.~173 as ``the precision on $\eta_{t+1}$ before observing $y_{t+1}$'' without clarifying that $\hat{h}_{t}$ is the precision after observing $y_{t}$ and before adding the shock $\delta_{t}$. 
} 
Bayesian updating implies
\begin{align}
\label{Bayes}
\eta_{t}|z^{t-1}\sim N\left(m_{t},\frac{1}{h_{t}}\right) = N\left(\frac{h_{t-1}m_{t-1}+h_{\epsilon}z_{t-1}}{h_{t-1}+h_{\epsilon}},\quad 1\middle/\frac{(h_{t-1}+h_{\epsilon})h_{\delta}}{h_{t-1}+h_{\epsilon}+h_{\delta}}
\right). 
\end{align}
The market sets the wage 
$w_{t}(y^{t-1})=m_{t}(z^{t-1})+a_{t}^*(y^{t-1})$, where $z^{t}=(z_{1},\ldots,z_{t})$. 
The common belief precision $h_{t}$ evolves deterministically, which is used to take it outside the expectations below. 

The manager's time-$t$ expectation $\mathbb{E}\left[w_{\tau}(y^{\tau-1})|y^{t-1}\right] $ of the future wage $w_{\tau}$, $\tau>t$ given $y^{t-1}$ and $\hat{a}_{t}$ is 
\begin{align}
\label{wage}
&\notag\mathbb{E}\left[w_{\tau}(y^{\tau-1})|y^{t-1}\right] 
=\mathbb{E}\left[\frac{h_{\tau-1}\frac{h_{\tau-2}m_{\tau-2}+h_{\epsilon}z_{\tau-2}}{h_{\tau-2}+h_{\epsilon}}+h_{\epsilon}z_{\tau-1}}{h_{\tau-1}+h_{\epsilon}}\middle|y^{t-1}\right]+\mathbb{E}\left[a_{\tau}^*(y^{\tau-1})|y^{t-1}\right]\\&\notag
=\mathbb{E}\left[\frac{h_{\tau-1}h_{\tau-2}m_{\tau-2}+h_{\tau-1}h_{\epsilon}z_{\tau-2}}{(h_{\tau-1}+h_{\epsilon})(h_{\tau-2}+h_{\epsilon})} +\frac{h_{\epsilon}z_{\tau-1}}{h_{\tau-1}+h_{\epsilon}}\middle|y^{t-1}\right]+\mathbb{E}\left[a_{\tau}^*(y^{\tau-1})|y^{t-1}\right]
\\&\notag
=\mathbb{E}\left[\frac{h_{\tau-1}h_{\tau-2}(h_{\tau-3}m_{\tau-3}+h_{\epsilon}z_{\tau-3})}{(h_{\tau-1}+h_{\epsilon})(h_{\tau-2}+h_{\epsilon})(h_{\tau-3}+h_{\epsilon})} + \frac{h_{\tau-1}h_{\epsilon}z_{\tau-2}}{(h_{\tau-1}+h_{\epsilon})(h_{\tau-2}+h_{\epsilon})} + \frac{h_{\epsilon}z_{\tau-1}}{h_{\tau-1}+h_{\epsilon}} +a_{\tau}^*(y^{\tau-1})|y^{t-1}\right]
\\&
=m_{t}\prod_{i=t}^{\tau-1}\frac{h_{i}}{h_{i}+h_{\epsilon}} +\frac{h_{\epsilon}}{h_{\tau-1}+h_{\epsilon}}\sum_{i=t}^{\tau-1}\mathbb{E}\left[z_{i}\middle|y^{t-1}\right]\prod_{j=i}^{\tau-2} \frac{h_{j+1}}{h_{j}+h_{\epsilon}} +\mathbb{E}\left[a_{\tau}^*(y^{\tau-1})|y^{t-1}\right], 
\end{align}
with the notational convention $\sum_{s=t+1}^{t}x_{s}=0$ and $\prod_{j=t}^{t-1}x_{s}=1$ for any $x_s$. 
The manager's expectation of the time-$t$ signal $z_{t}$ is $\mathbb{E}\left[\eta_{t}+\epsilon_{t}+\hat{a}_{t}-a_{t}^*(y^{t-1})\middle|y^{t-1}\right] =m_{t}+\hat{a}_{t}-a_{t}^*(y^{t-1})$. When forming expectations about future signals $z_{i}$, $i>t$, the manager expects her future selves to follow the equilibrium strategy, therefore  $\mathbb{E}\left[z_{i}\middle|y^{t-1}\right] =\mathbb{E}\left[\eta_{i}+\epsilon_{i}+a_{i}^*(y^{i-1})-a_{i}^*(y^{i-1})\middle|y^{t-1}\right]=m_{t}$. 
Substituting the expected signals into~(\ref{wage}) results in 
\begin{align}
\label{wage2}
\mathbb{E}\left[w_{\tau}(y^{\tau-1})|y^{t-1}\right]=
& m_{t}\prod_{j=t}^{\tau-1}\frac{h_{j}}{h_{j}+h_{\epsilon}} +\frac{h_{\epsilon}}{h_{\tau-1}+h_{\epsilon}}[m_{t}+\hat{a}_{t}-a_{t}^*(y^{t-1})]\prod_{j=t}^{\tau-2} \frac{h_{j+1}}{h_{j}+h_{\epsilon}} 
\\&\notag
+\frac{h_{\epsilon}}{h_{\tau-1}+h_{\epsilon}}m_{t}\sum_{i=t{\color{red}+1}}^{\tau-1}\prod_{j=i}^{\tau-2} \frac{h_{j+1}}{h_{j}+h_{\epsilon}} +\mathbb{E}\left[a_{\tau}^*(y^{\tau-1})|y^{t-1}\right]. 
\end{align}
Substituting~(\ref{wage2}) into~(\ref{equilibrium}) 
yields the manager's objective function 
\begin{align}
\label{objective}
&\notag m_{t}+a^{*}_{t}(y^{t-1})-g(\hat{a}_{t}) + m_{t}\sum_{\tau=t+1}^{\infty}\beta^{\tau-t}\prod_{j=t}^{\tau-1}\frac{h_{j}}{h_{j}+h_{\epsilon}} +[\hat{a}_{t}-a_{t}^*(y^{t-1})]\sum_{\tau=t+1}^{\infty}\beta^{\tau-t}\frac{h_{\epsilon}}{h_{\tau-1}+h_{\epsilon}}\prod_{j=t}^{\tau-2} \frac{h_{j+1}}{h_{j}+h_{\epsilon}} 
\\& +m_{t}\sum_{\tau=t+1}^{\infty}\beta^{\tau-t}\frac{h_{\epsilon}}{h_{\tau-1}+h_{\epsilon}}\sum_{i={\color{red}t}}^{\tau-1}\prod_{j=i}^{\tau-2} \frac{h_{j+1}}{h_{j}+h_{\epsilon}}  +\sum_{\tau=t+1}^{\infty}\beta^{\tau-t}\mathbb{E}\left[a_{\tau}^*(y^{\tau-1}) -g(a_{\tau}^{*}(y^{\tau-1}))|y^{t-1}\right]
\end{align}
and taking the first order condition (FOC) w.r.t.\ $\hat{a}_{t}$ results in 
\begin{align}
\label{foc}
\sum_{s=t+1}^{\infty}\beta^{s-t}\frac{h_{\epsilon}}{h_{s-1}+h_{\epsilon}}\prod_{j=t}^{s-2} \frac{h_{j+1}}{h_{j}+h_{\epsilon}}-g'(\hat{a}_{t})=0. 
\end{align}
The manager's marginal benefit of $\hat{a}_{t}$ is the discounted sum of the responses of future wages to $\hat{a}_{t}$. Because the wage is paid in advance, period-$t$ effort does not affect the period-$t$ wage. The SOC holds, because $g$ is convex. 
When $h_{\delta}=\infty$, the FOC in 
\cite{holmstrom1999} is Equation~(10), reproduced in~(\ref{holmstromfoc1}):
\begin{align}
\label{holmstromfoc1}
\tag{H10}
\sum_{s=t}^{\infty}\beta^{s-t}\frac{h_{\epsilon}}{h_{s}}=g'(a_{t}^{*}).
\end{align} 
It corresponds to the current paper's FOC~(\ref{foc}) with $h_{\delta}=\infty$, in which case $h_{s-1}+h_{\epsilon}=h_{s}$, but \citeauthor{holmstrom1999} erroneously starts the sum from index $t$, not $t+1$. 
\citeauthor{holmstrom1999}'s Equation~(21), reproduced below in~(\ref{holmstromfoc2}), corresponds to the current paper's~(\ref{foc}) for general $h_{\delta}$ and starts the sum correctly from $t+1$, but has a different error. 
\cite{holmstrom1999} defines $\mu_{t}:=\frac{h_{t}}{h_{t}+h_{\epsilon}}$, and his FOC is 
\begin{align}
\label{holmstromfoc2}
\tag{H21}
(1-\mu_{t})\sum_{s=t+1}^{\infty}\beta^{s-t}\prod_{i=t+1}^{s}\mu_{i} =g'(a_{t}^{*}). 
\end{align}
Taking $h_{\delta}=\infty$ (so that $h_{i}+h_{\epsilon}=h_{i+1}$), 
the LHS 
in~(\ref{holmstromfoc2}) becomes 
$\frac{h_{\epsilon}}{h_{t}+h_{\epsilon}}\sum_{s=t+1}^{\infty}\beta^{s-t}\prod_{i=t+1}^{s}\frac{h_{i}}{h_{i}+h_{\epsilon}}
=\sum_{s=t+1}^{\infty}\beta^{s-t}\frac{h_{\epsilon}}{h_{t+1}}\prod_{i=t+1}^{s}\frac{h_{i}}{h_{i+1}}
=\sum_{s=t+1}^{\infty}\beta^{s-t}\frac{h_{\epsilon}}{h_{s+1}}$, which is inconsistent with~(\ref{holmstromfoc1}) where the subscript of $h_{s+1}$ is $s$. 
To make~(\ref{holmstromfoc2}) consistent with~(\ref{holmstromfoc1}) when $h_{\delta}=\infty$ and the sum in~(\ref{holmstromfoc1}) has been corrected to start at $t+1$, the product in~(\ref{holmstromfoc2}) should end at $s-1$, not $s$. Thus~(\ref{holmstromfoc2}) should be written 
\begin{align}
\label{holmstromfoc2mod}
(1-\mu_{t})\sum_{s=t+1}^{\infty}\beta^{s-t}\prod_{i=t+1}^{s-1}\mu_{i} =g'(a_{t}^{*}).
\end{align} 
The LHS of~(\ref{holmstromfoc2mod}) is 
$\sum_{s=t+1}^{\infty}\beta^{s-t}\frac{h_{\epsilon}}{h_{t}+h_{\epsilon}}\prod_{i=t+1}^{s-1}\frac{h_{i}}{h_{i}+h_{\epsilon}} 
=\sum_{s=t+1}^{\infty}\beta^{s-t}\frac{h_{\epsilon}}{h_{s-1}+h_{\epsilon}}\prod_{i=t+1}^{s-1}\frac{h_{i}}{h_{i-1}+h_{\epsilon}}$, same as the marginal benefit in~(\ref{foc}). 
The intuition for the correct FOC is that increasing the manager's labor $\hat{a}_t$ by one unit increases the market's mean belief $m_{t+1}$ next period by $\frac{h_{\epsilon}}{h_{t}+h_{\epsilon}}$ units, but given $m_{t+1}$, does not directly affect any $m_{\tau}$, $\tau\neq t+1$. For any $n\in\mathbb{N}$, increasing $m_{n}$ by one unit raises $m_{n+1}$ by $\frac{h_{n}}{h_{n}+h_{\epsilon}}$ units, but does not directly affect any $m_{\tau}$, $\tau\notin\set{n,n+1}$. Therefore increasing $\hat{a}_{t}$ by one unit raises $m_{t+k}$, $k\geq 1$ by $\frac{h_{\epsilon}}{h_{t}+h_{\epsilon}}\prod_{i=1}^{k-1}\frac{h_{t+i}}{h_{t+i}+h_{\epsilon}}$ units. 
The expected wage $\mathbb{E}[w_{t+k}|y^{t-1}]$ increases one for one with $m_{t+k}$, because $a_{n}^*$ is independent of $m_{\ell}$ for any $n,\ell$, thus there is no multiplier effect from $\hat{a}_{t}$ to $m_{n+k}$ via $a_{n}^*$. 
A unit increase in $w_{t+k}$ is worth $\beta^{k}$ at time $t$.

\subsection{The stationary case}
\label{sec:stationary}

A \emph{stationary equilibrium} requires $h_{\delta}<\infty$ and features a constant belief precision $h_{t+1}=h_t=:h$. 
In terms of $\mu:=\frac{h}{h+h_{\epsilon}}$, the steady state marginal benefit of the manager's action (the LHS of~(\ref{holmstromfoc2mod})) is 
\begin{align}
\label{steadystate}
(1-\mu)\sum_{s=1}^{\infty}\beta^{s}\prod_{i=1}^{s-1}\mu 
=(1-\mu)\sum_{s=1}^{\infty}\beta^{s}\mu^{s-1}
=\frac{\beta(1-\mu)}{1-\beta\mu}.
\end{align} 
This matches \cite{holmstrom1999} Equation~(22), reproduced below in~(\ref{steadylabor}), but is not derived from~(\ref{holmstromfoc2}). The steady state version of~(\ref{holmstromfoc2}) is 
instead $(1-\mu)\sum_{s=1}^{\infty}\beta^{s}\prod_{i=1}^{s}\mu 
=\frac{\beta\mu(1-\mu)}{1-\beta\mu}$, which has different comparative statics from those in \cite{holmstrom1999} Proposition~1. 


For the stationary labor supply $a^{*}$ to satisfy \citeauthor{holmstrom1999}'s Equation~(22) 
\begin{align}
\label{steadylabor}
\tag{H22}
\frac{\beta(1-\mu)}{1-\beta\mu} =g'(a^{*}), 
\end{align} 
the assumptions $g\in C^1$, $g'(0)=0$ and $\lim_{\hat{a}\rightarrow\infty}g'(\hat{a})>1$ are sufficient. 
\cite{holmstrom1999} p.~174 states that $a^{*}$
is close to zero when $\beta<1$ and $h_{\delta}$ is large relative to $h_{\epsilon}$, which is true iff 
the additional\footnote{
Recall that \cite{holmstrom1999} only assumes that $g$ is increasing and convex. 
} assumption that $g'(a)=0\Rightarrow a=0$ holds. 
If this assumption is violated, i.e.\ if $g'(a)=0$ for all $a\in[0,k]$ for some $k>0$, then as the marginal benefit of labor (the LHS of~(\ref{steadylabor})) converges to zero, the manager's labor supply $a^{*}$ converges to $k$, not zero. 

\cite{holmstrom1999} p.~174 compares the manager's labor supply under $h_{\delta}=\infty$ and $h_{\delta}<\infty$, both when $\beta<1$ and when $\beta=1$. 
The comparison is indeterminate if $\beta=1$, because in this case, one of the summands in the manager's utility~(\ref{objective}) is infinite\footnote{
Proof:  $\frac{h_{j+1}}{h_{j}+h_{\epsilon}}\geq\min\set{\frac{h_{t+1}}{h_{t}+h_{\epsilon}},\frac{h}{h+h_{\epsilon}}} =:\iota>0$ uniformly in $j\geq t$, so $\sum_{i=t}^{\tau-1}\prod_{j=i}^{\tau-2}\frac{h_{j+1}}{h_{j}+h_{\epsilon}} \geq \frac{1-\iota^{\tau-t}}{1-\iota} > 1$ and $\sum_{\tau=t+1}^{\infty}\frac{h_{\epsilon}}{h_{\tau-1}+h_{\epsilon}}\sum_{i=t}^{\tau-1}\prod_{j=i}^{\tau-2}\frac{h_{j+1}}{h_{j}+h_{\epsilon}} =\infty$. 
},
so the maximizer of~(\ref{objective}) is undefined. If $h_{\delta}<\infty$ and $\lim_{\hat{a}\rightarrow\infty}g'(\hat{a})=\infty$, then the FOC~(\ref{holmstromfoc2mod}) still has a finite solution, which may be defined as the manager's choice of labor when~(\ref{objective}) is infinite. In that case the rest of the analysis remains valid. 

If $\beta=1$ and $h_{\delta}=\infty$, then the marginal benefit of labor in the FOC~(\ref{holmstromfoc2mod}) diverges, so the solution of the FOC is $a_{t}^{*}=\infty$. This infinite labor presents a problem for Bayesian updating and by extension the derivation of the FOC itself. 
If the market expects $a_{t}^*(y^{t-1})=\infty$, then any $\hat{a}_{t}<\infty$ leads to $y_{t}<\infty$, which is off the equilibrium path, so Bayes' rule does not apply. One way to resolve the updating problem is to use belief threats to deter the manager from deviating, e.g.\ set $m_{t+1}=-\infty$ after $y_{t}<\infty$. 
If the manager chooses $\hat{a}_{t}=\infty$, then the output is $y_{t}=\infty$ for any $\eta_{t}$, thus uninformative about the type. By Bayes' rule, the market's mean belief is $m_{t+1}=m_{t}$ after $y_{t}=\infty$.

\subsection{Transient effects}
\label{sec:transient}

On p.~174, \cite{holmstrom1999} defines $b_{s}(\mu_1):=(1-\mu_1)\mu_{2}\cdots \mu_{s}$ without clarifying whether the $\mu_{i}$, $i>1$ are independent variables or 
the functions of $\mu_1$ recursively defined in \citeauthor{holmstrom1999}'s Equation~(17): 
\begin{align} 
\label{H17}
\tag{H17}
\mu_{t+1}=\frac{1}{2+h_{\epsilon}/h_{\delta}-\mu_{t}}. 
\end{align}
This distinction becomes important in the equation 
\begin{align}
\label{transient}
b_{s+1}(\mu_1) =\frac{1-\mu_1}{1-\mu_2}\mu_2b_{s}(\mu_2) =\frac{1-\mu_1}{1+r-\mu_1}b_{s}(\mu_2), 
\end{align}
which is unlabeled 
in \cite{holmstrom1999}. 
If all $\mu_{i}$ are treated as independent variables, then $b_{s}(\mu_2)=(1-\mu_2)\mu_2\prod_{i=3}^{s}\mu_{i}$, but if each $\mu_{i}$ is understood as a function of the argument of $b_{s}$, then $b_{s}(\mu_2)=(1-\mu_2)\prod_{i=3}^{s}\mu_{i}$, because each $\mu_{i}$ is then obtained from $\mu_2$ by $i-2$ iterations of~(\ref{H17}). 
In either case, the first equality in~(\ref{transient}) fails in general. 
If all $\mu_{i}$ are independent variables and $\mu_{2}\neq \mu_{s+1}$, then 
\begin{align*}
b_{s+1}(\mu_1) =(1-\mu_1)\mu_{s+1}\prod_{i=2}^{s}\mu_{i} \neq \frac{1-\mu_1}{1-\mu_2}\mu_2b_{s}(\mu_2) =(1-\mu_1)\mu_2^2\prod_{i=3}^{s}\mu_{i} . 
\end{align*}
If each $\mu_{i}$ is treated as a function of the argument of $b_{s}$, and $\mu_{s+1}\neq 1$, then 
\begin{align*}
(1-\mu_1)\mu_{s+1}\prod_{i=2}^{s}\mu_{i} \neq \frac{1-\mu_1}{1-\mu_2}\mu_2b_{s}(\mu_2) =(1-\mu_1)\prod_{i=2}^{s}\mu_{i}. 
\end{align*}
From now on, treat each $\mu_{i}$ as a function of the argument of $b_{s}$. 
A modification to~(\ref{transient}) that makes the equalities in it hold is to replace $b_{s}$ with $b_{s+1}$, obtaining $b_{s+1}(\mu_1) =\frac{1-\mu_1}{1-\mu_2}\mu_2b_{s+1}(\mu_2) =\frac{1-\mu_1}{1+r-\mu_1}b_{s+1}(\mu_2)$. Another way to make~(\ref{transient}) valid is to define $b_{s}(\mu_2)$ as $(1-\mu_{2})\mu_{3}\cdots\mu_{s+1}$, which suggests the general definition 
\begin{align}
\label{bsredef}
b_{s}(\mu_t) :=(1-\mu_{t})\prod_{i=t+1}^{t+s-1}\mu_{i}\quad \text{for any } s\geq 1\text{ and }1\leq t\leq s.
\end{align} 
Either way to make~(\ref{transient}) hold ensures that the inductive proof on p.~174 of \cite{holmstrom1999} is correct, i.e.\ that $\gamma_1:= (1-\mu_{1})\sum_{s=2}^{\infty}\beta^{s-1}\prod_{i=2}^{s}\mu_{i}$ (the LHS of~(\ref{holmstromfoc2}) at $t=1$) decreases in $\mu_1$. 
The same proof shows that the LHS of~(\ref{holmstromfoc2mod}) at $t=1$ decreases in $\mu_{1}$. 
Using~(\ref{bsredef}) in~(\ref{transient}) and replacing $\mu_1,\mu_2$ with $\mu_{t},\mu_{t+1}$ respectively shows that for any $t$, the LHS-s of~(\ref{holmstromfoc2}) and~(\ref{holmstromfoc2mod}) decrease in $\mu_{t}$.  This proves Proposition~2 in \cite{holmstrom1999}. 

If the definition of $b_{s}$ is not altered to~(\ref{bsredef}), but remains $b_{s}(\mu_t)=(1-\mu_t)\prod_{i=t+1}^{s}\mu_{i}$, then to prove that for any $t$, the LHS of~(\ref{holmstromfoc2}) decreases in $\mu_{t}$, the equation~(\ref{transient}) should be modified separately for each $t$. Specifically, $\mu_{1}$ and $\mu_{2}$ should be replaced with $\mu_{t}$ and $\mu_{t+1}$ respectively, and both $b_{s+1}$ and $b_{s}$ should be replaced with $b_{s+t}$. 

\bibliographystyle{ecta}
\bibliography{teooriaPaberid} 
\end{document}